\documentclass[seceqn,fleqn]{elsart}
\usepackage{amsmath,amssymb}
\usepackage{xspace}
\usepackage{epsfig}

\newcommand {\version}{v4}                     
\newcommand {\versiondate}{November 2, 2005}   

\renewcommand{\vec}[1]{\boldsymbol{\mathrm{#1}}}

\DeclareMathOperator{\sgn}{sgn}
\DeclareMathOperator{\arcsinh}{arcsinh}

\newcommand{\diag}{\ensuremath{\mathrm{diag}}}  
\newcommand{\defeq}{\ensuremath{\equiv}}        
\newcommand{\qefed}{\ensuremath{\equiv}}        
\newcommand{\id}{\ensuremath{\mathrm{d}}}       
\newcommand{\ii}{\mathrm{i}}                    
\newcommand{\unit}[1]{\;\mathrm{#1}}            

\newcommand{\Cerenkov}{\v{C}erenkov\xspace}     
\newcommand{\DS}[1]{$#1$\xspace}                
\newcommand{\action}{\mathcal{S}}               

\newcommand{\sqroot}[1]{\sqrt{\smash[b]{\mathstrut}#1}}

\newcommand{\abs}[1]{|{#1}|}                    

\newcommand{\BigO}{\mathrm{O}}                  

\newcommand{\Meff}[1][]{\ensuremath{M_{\text{eff}#1}}}
\newcommand{\Meffs}[1][]{\ensuremath{M_{\text{eff}#1}^2}}

\newcommand{\plus}{\oplus}                      
\newcommand{\minus}{\ominus}                    

\newcommand{\Mpsi}{M_\psi} 
\newcommand{\Mphi}{M_\phi} 
\newcommand{\boxpower}{l}  

\newcommand{\vx}{\vec \xi}
\newcommand{\vy}{\vec \eta}
\newcommand{\vz}{\vec \zeta}

\newcommand{\vk}{\vec k}
\newcommand{\vq}{\vec q}
\newcommand{\vf}{\vec f}
\newcommand{\vb}{\vec b}
\newcommand{\vv}{\vec v}

\newcommand{\kp}{k_\parallel}
\newcommand{\kpabs}{|k_\parallel|}
\newcommand{\kpa}[1]{k_{\parallel,#1}}
\newcommand{\ko}{k_\perp}
\newcommand{\koa}[1]{k_{\perp,#1}}
\newcommand{\vko}{\vec k_\perp}
\newcommand{\vkoa}[1]{\vec k_{\perp,#1}}

\newcommand{\qp}{q_\parallel}
\newcommand{\qpabs}{|q_\parallel|}
\newcommand{\qo}{q_\perp}
\newcommand{\vqo}{\vec q_\perp}
\newcommand{\kqq}{(kqq)}             

\newcommand{\omegap}{\omega_\parallel}
\newcommand{\omegapa}[1]{\omega_{\parallel,#1}}

\begin{document}

\begin{frontmatter}
\noindent Nucl. Phys. B 734 (2006) 1
\hfill  hep-th/0508074 (\version)\newline  

\title{Vacuum Cherenkov radiation and photon triple-splitting
in a Lorentz-noninvariant extension of quantum electrodynamics}
\author{C. Kaufhold}\ead{kaufhold@particle.uni-karlsruhe.de},
\author{F.R. Klinkhamer}
\ead{frans.klinkhamer@physik.uni-karlsruhe.de}
\address{Institute for Theoretical Physics, University of
 Karlsruhe (TH), 76128 Karlsruhe, Germany}

\begin{abstract}
We consider a \DS{CPT}--noninvariant scalar model and a
modified version of quantum electrodynamics with
an additional photonic Chern--Simons-like term in the action.
In both cases, the Lorentz violation
traces back to a spacelike background vector.
The effects of the modified field equations and dispersion relations on
the kinematics and dynamics of decay processes are discussed,
first for the simple scalar model and then for modified quantum electrodynamics.
The decay widths for electron Cherenkov radiation
in modified quantum electrodynamics and for photon triple-splitting
in the corresponding low-energy effective theory
are obtained to lowest order in the electromagnetic coupling constant.
A conjecture for the high-energy limit of the photon-triple-splitting
decay width at tree level is also presented.
\end{abstract}


\date{\versiondate}
\end{frontmatter}

\section{Introduction}\label{sec:introduction}

The Lorentz-noninvariant Maxwell--Chern--Simons (MCS) model in four
spacetime dimensions has been extensively studied over the last years
\mbox{\cite{CFJ:MCS,CK:Lorentz,AK:causality}.}
With suitable interactions added, the model allows for photon
triple-splitting ($\gamma\to\gamma\gamma\gamma$), but the decay
width of the photon has only been calculated for one special case
\cite{AK:photon-decay}.
Another interesting process is vacuum Cherenkov radiation, with an
electron emitting an MCS photon ($e^- \to e^- \gamma$),
for which there exists an order of magnitude estimate of
the decay probability \cite{LP:Cherenkov}.

In order to improve our understanding,
we consider in the present article simple
scalar models where Lorentz symmetry is broken by
a (purely) spacelike background vector, and focus on the
effects in decay processes. We, then, turn to two- and three-particle
decays in a modified version of quantum electrodynamics (QED),
which has an anisotropic photonic Chern--Simons-like term
\cite{CFJ:MCS,CK:Lorentz,AK:causality}
added to the standard Maxwell--Dirac action of QED \cite{JauchRohrlich}.
First,  we  calculate the vacuum Cherenkov radiation of a moving
electron (i.e., moving with respect to the preferred frame from the
Chern--Simons-like term).
Second, we obtain the general result for
photon triple-splitting in the low-energy effective theory with electrons
integrated out (specifically, the photon model with the Euler--Heisenberg
quartic interaction term added to the quadratic MCS terms).
Third, we discuss the possible high-energy behavior of
the photon-triple-splitting decay width at tree level,
which may be compared to the behavior of
the exact tree-level result for vacuum Cherenkov radiation.

The scope of the present article has been restricted to a relatively simple
and well-understood model, as our goal is to perform a \emph{complete}
calculation of certain two- and three-particle decays, at least at tree level.
In the pure photon sector, there are only two
possible bi-linear Lorentz-violating terms \cite{CK:Lorentz},
one of which, the \DS{CPT}--odd Chern--Simons-like term, we study in
detail. The other possible term, which is \DS{CPT}--even,
will be discussed briefly at the end of the article.

This article is organized as follows. In Section \ref{sec:lorentz},
we present the scalar and photon models considered
and introduce the concept of effective mass square.
In Section \ref{sec:decay}, we define the
decay width in a Lorentz-violating theory and,
in Section \ref{sec:scalar},  study a few simple decay processes for scalars.
In Section \ref{sec:cherenkov}, we give results
for Cherenkov radiation in modified QED with a spacelike
Chern--Simons-like term and, in Section \ref{sec:photon}, for
photon  triple-splitting in the corresponding low-energy effective theory
(details of the calculation are relegated
to the appendices).
In Section \ref{sec:discussion}, we place our results in a
larger context and speculate on photon triple-splitting from
the \DS{CPT}--even Lorentz-violating term mentioned above.

The present article is rather technical and the reader who is only
interested in the main results may skip ahead to
Sections~\ref{sec:cherenkov} and \ref{sec:photon}, while referring
back to Sections~\ref{sec:lorentz} and \ref{sec:decay} for
the necessary definitions.

\section{Lorentz violation and effective mass squares}
\label{sec:lorentz}

\subsection{Framework and conventions}

We work in four-dimensional Minkowski spacetime, and postulate
translation invariance which implies energy-momentum conservation.
The Lorentz-symmetry breaking is implemented by the presence of
constant background tensors in the action, as in the
Standard Model extension of Ref.~\cite{CK:Lorentz}.  We are mainly
interested in bi-linear breaking terms which lead to  modified
dispersion relations and  new definitions of the free particle states.
As will be seen later, such modifications are more interesting than
having just one particular type of Lorentz-violating interaction,
since the changed kinematics appears in all kinds of physical processes.

In our examples, Lorentz invariance is broken by a real background
four-vector
\begin{equation}
\zeta^\mu = (\zeta^0, \vz),\qquad\zeta^\mu \qefed m \, \widehat\zeta^{\,\mu},
\end{equation}
where we have split $\zeta^\mu$ into a dimensionless unit vector
$\widehat\zeta^{\,\mu}=(\widehat\zeta^{\,0},\widehat\vz)$
characterizing the Lorentz-breaking direction in spacetime (normalization
$\widehat\zeta^{\,\mu} \widehat\zeta_\mu=\pm1$) and a mass scale $m > 0$
setting its length. If $\zeta^\mu$ is spacelike, there is a
preferred spatial direction in coordinate frames with $\zeta^0 =0$.
For timelike $\zeta^\mu$, there is rotational invariance in frames with
$\vz=0$. We do not consider the ``lightlike'' case ($\zeta^\mu \zeta_\mu =0$),
which can be viewed as being of measure zero.

In the purely spacelike case ($\zeta^0 =0$ and $|\widehat\vz|=1$),
we will use the following notations:
\begin{equation}
\kp \defeq \vk\cdot\widehat\vz\text,\quad \vko \defeq \vk - \kp\, \widehat\vz,
\quad \ko\defeq\abs{\vko}\text,
\end{equation}
for an arbitrary three-vector $\vk$.
Throughout, we employ the Minkowski metric $(\eta_{\mu\nu})=\diag(+1,-1,-1,-1)$,
take $\epsilon_{0123}=1$, and set $\hbar=c=1$ (except when stated otherwise).

In the main part of this article, we assume that $\zeta^\mu$ is
spacelike and choose a particular
coordinate frame so that  $\zeta^\mu$ is purely spacelike.
The calculations are simplified in such a frame
(loosely called the ``purely spacelike frame''),
because Lorentz invariance with respect
to boosts in directions orthogonal to $\vz$ is preserved, together
with rotational invariance around $\vz$. In
Section~\ref{sec:lifetime}, we discuss how to interpret these theories in a
general frame.

\subsection{Maxwell--Chern--Simons model}

As mentioned in the Introduction,
we start from a noninteracting photon model which results from
adding a \DS{CPT}--odd Abelian Chern--Simons-like term
\cite{CFJ:MCS,CK:Lorentz,AK:causality} to the standard Maxwell term
\cite{JauchRohrlich}. The model action is then given by
\begin{equation}\label{eq:MCS}
\action_\text{MCS}= \action_\text{M}  + \action_\text{CS} \,\text,
\end{equation}
with the following Maxwell and Chern--Simons-like  terms:
\begin{eqnarray}
\action_\text{M} &=& \int_{\mathbb{R}^4} \mathrm{d}^4 x
\left(\, -\frac14\;F_{\mu\nu}F^{\mu\nu}\, \right)\text,
\label{eq:Mterm}\\[2mm]
\action_\text{CS} &=& \int_{\mathbb{R}^4} \mathrm{d}^4 x
\left(\,\frac14\;\, m\,\epsilon_{\mu\nu\rho\sigma}\,\widehat\zeta^{\,\mu}\,A^\nu\,
F^{\rho\sigma}\, \right)\text,
\label{eq:CSterm}
\end{eqnarray}
where $A_\mu(x)$ is the gauge field and $F_{\mu\nu}(x)=\partial_\mu
A_\nu(x)-\partial_\nu A_\mu(x)$ the  field strength.
The action \eqref{eq:CSterm} is gauge invariant, provided
the field strength vanishes rapidly enough at infinity.

The extra term \eqref{eq:CSterm} may, for example,
be induced by the \DS{CPT} anomaly of
chiral gauge theories over a topologically nontrivial spacetime
manifold \cite{K:CPT,KS:CPT,K:CPT-review2005}. From astrophysical
bounds  \cite{CFJ:MCS,CK:Lorentz,WPC:97}, the mass
parameter $m$ is known to be very small, at least for the
regular photon and the present universe.
The relatively simple model \eqref{eq:MCS} may,
however, have other applications and the intention of
the present article is purely theoretical.

If the background vector $\zeta^\mu$ is spacelike, quantization is
possible in frames where this  vector is purely spacelike
\cite{AK:causality}. In these frames,
the invariance under time reversal  \DS{T} is broken,
while  charge conjugation \DS{C} and parity reflection \DS{P}
remain symmetries; cf. Refs.~\cite{K:CPT-review2005,Sakurai,CK:CPT}.
For timelike $\zeta^\mu$, a quantization does not appear to be
possible \cite{AK:causality}.

The field equations of the model \eqref{eq:MCS} are
\begin{equation}\label{eq:field-equation}
\big(\Box \eta^{\mu\nu} - \partial^\mu\partial^\nu -  \,m\,
 \epsilon^{\mu\nu\rho\sigma}\widehat\zeta_{\rho}\, \partial_\sigma\big)\,
A_\nu=0\text.
\end{equation}
From a plane-wave \emph{Ansatz}, one obtains the dispersion relation
for the momentum four-vector $k^\mu$,
\begin{equation}\label{eq:dispersion}
(k^\mu k_\mu)^2+ (k^\mu k_\mu) (\zeta^\nu\zeta_\nu)- (k^\mu
  \zeta_\mu)^2=0\text,
\end{equation}
with $\zeta^\nu \equiv m \, \widehat\zeta^{\,\nu}$.
This relation gives two different propagation modes, called $\plus$ and
$\minus$ in the following, according to the form the dispersion
relation takes in a purely spacelike frame,
\begin{equation}
\omega_\pm(\vk)=\sqrt{\big(\omegapa{\pm}(\kp)\big)^2+\ko^2}\,\text,
\label{eq:MCS-omega}
\end{equation}
with  ``parallel energies''
\begin{align}
\omegapa{\pm}(\kp) & \defeq
\sqrt{\kp^2+m^2/4}\, \pm\,  m/2 \sim \kpabs \pm m/2
\label{eq:MCS-omegaparallel}\text,
\end{align}
for $\kpabs \gg m > 0$.

The $\plus$ mode has timelike four-momentum and a mass gap.
The $\minus$ mode has spacelike four-momentum and is gapless.
But, even though certain photon momenta can be spacelike, the theory is
still causal \cite{AK:causality}. In Appendix~\ref{ap:polarizations},
we present the modified polarization vectors of the photon,
together with certain useful relations. For further details,
see, in particular, Section~2 of Ref.~\cite{AK:photon-decay}.

\subsection{Lorentz-noninvariant scalar model}

The Lorentz-noninvariant photon model of the previous subsection is
already quite complicated and, as a simpler example, we consider a
complex scalar field with the following free action:
\begin{equation}\label{eq:scalar}
\action_\text{scalar}=\int_{\mathbb{R}^4} \id^4 x \left( \partial_\mu
\bar\phi \,
 \partial^\mu \phi
 +  \, \ii \,  m\,\bar\phi \, \widehat\zeta^\mu \partial_\mu \phi
 -  \, \ii \,  m\,    \phi \, \widehat\zeta^\mu \partial_\mu \bar\phi
 -  \Mphi^2 \,\bar\phi \phi \right)\text,
\end{equation}
with $m$ and  $\Mphi$ taken positive.
A spacetime-dependent phase redefinition of the fields could
eliminate the Lorentz-violating term in the free model; cf.
Refs. \cite{CK:CPT,A:scalar}. Here, however,  we assume that
the interactions considered  do not allow these phase redefinitions
(see Section~\ref{sec:scalar}).

In this scalar model, the conventional discrete symmetries
\DS{C}, \DS{P}, and \DS{T}  are broken \cite{Sakurai}. The combinations
\DS{PT}, \DS{CP}, \DS{CT} are conserved, but not \DS{CPT}.

The action \eqref{eq:scalar} yields the field equation
\begin{equation}
\Box \phi - 2\, \ii \, m \, \widehat\zeta^{\,\mu} \partial_\mu \phi
+ \Mphi^2 \, \phi = 0\text,
\end{equation}
and the complex-conjugate equation for $\bar\phi$.
The dispersion relation for the momentum vector $k^\mu$ is then given by:
\begin{equation}\label{eq:D-scalar}
k^\mu k_\mu \pm 2 m \, \widehat\zeta^{\,\mu} k_\mu - \Mphi^2 = 0\text,
\end{equation}
with the upper sign for $\phi$ and the lower for $\bar\phi$.

Solving the quadratic  $k^0$ equation \eqref{eq:D-scalar}, one has for
the positive energy branch,
\begin{align}\label{eq:omegascalar}
\omega(\vk)&=\sqrt{|\vk|^2 \pm 2 \vk\cdot\vz  + (\zeta^0)^2 + \Mphi^2}\,
\mp  \zeta^0 \text,
\end{align}
with $\zeta^\mu \equiv m \, \widehat\zeta^{\,\mu}$.
For the purely spacelike case, the energy reduces to
\begin{equation}
\omega(\vk)=\sqrt{|\vk|^2 \pm 2 \vk \cdot \vz +
  \Mphi^2}=\sqrt{\omegap(\kp)^2+\ko^2}\text,
\end{equation}
with
\begin{align}
\omegap(\kp)&\defeq\sqrt{\kp^2\pm 2 m \kp + \Mphi^2}
\sim  \kpabs \pm m \, \sgn{\kp}\text,
\label{eq:scalar-omegaparallel}
\end{align}
for $\kpabs \gg \max(m,\Mphi)$.
The asymptotic parallel energy \eqref{eq:scalar-omegaparallel}
of the scalar model has the same structure as the result
\eqref{eq:MCS-omegaparallel} of the MCS model, apart from
the sign of the mass term (remember that \DS{C}, \DS{P}, and \DS{T} are
broken in the scalar model, whereas
only \DS{T} is broken in the purely spacelike MCS model).

For $\Mphi^2+\zeta^\mu \zeta_\mu \geq 0$, the group
velocity~$\abs{\partial\omega/\partial \vk}$ is less or equal to $1$;
cf. Ref.~\cite{A:scalar}. Microcausality (locality) can then be verified
by a calculation analogous to the Lorentz-invariant case.
Recall  that locality \cite{JauchRohrlich} requires, in particular,
a vanishing commutator
$[ \phi(x), \bar\phi(y)]$ for spacelike separation  $x^\mu-y^\mu$.

The Lorentz-violating terms in \eqref{eq:MCS} and  \eqref{eq:scalar}
have a similar structure: each term is bi-linear in the fields,
has a single derivative, and couples to the background vector.
The models also have similar dispersion relations $\omega(\vk)$,
as long as the  scalar mass parameter $\Mphi$
does not dominate, $\Mphi \ll \abs{\vk}$.
A photon mass term in the MCS model is, of course, forbidden by gauge invariance.

\subsection{Effective mass squares}\label{sec:masses}

One way to look at modified dispersion relations is to think
of the norm square of the energy-momentum four-vector as the ``effective
mass square'' of the particle with corresponding three-momentum,
\begin{equation}\label{eq:Meffs}
\Meffs(\vk) \equiv \omega(\vk)^2-|\vk|^2 = k^\mu k_\mu\text.
\end{equation}
This effective mass square is associated with a specific mode
and may depend on the coordinate frame used.\footnote{The concept of
effective mass square appears already in Refs.~\cite{OS,CG:Lorentz} and,
more recently, has been used in Ref.~\cite{GNY:photon-splitting}.}
For timelike momentum, on the one hand, the effective mass square
is the energy square in the frame
with zero three-momentum. For spacelike momentum, on the other hand,
the effective mass square is negative and there is no
frame with vanishing three-momentum.

The effective mass square of the  photon from
the purely spacelike MCS model \eqref{eq:MCS} is given by
\begin{align}\label{eq:Meffs-MCS}
\Meffs[,\,\text{MCS}\,\pm](\vk) &=\pm\, m \, \omegapa{\pm}(\kp) \
= \pm m\, \kpabs + m^2/2 + \BigO\big(m^3/\kpabs\big)\text,
\end{align}
for $\kpabs \gg m$. The absolute value of this effective mass square grows
with the momentum component in the preferred direction, but is still
suppressed relative to the energy square,
\begin{equation}
\frac{|\Meffs[,\,\text{MCS}\,\pm]|}{\omega_\pm^2} \leq
\frac{|\Meffs[,\,\text{MCS}\,\pm]|}{\omegapa{\pm}^2}
\sim \frac{m}{\kpabs}\,\text,
\end{equation}
for $\kpabs\gg m$.

The effective mass square of the scalar model \eqref{eq:scalar}
in the purely spacelike frame is given by
\begin{equation}\label{eq:Meffs-scalar}
\Meffs[,\,\text{scalar}](\vk)=\pm 2\, m\, \kp + \Mphi^2\text,
\end{equation}
with the upper sign for $\phi$ and the lower for $\bar\phi$.
The (anti-)scalar effective mass square from Eq.~\eqref{eq:Meffs-scalar}
has the same structure as the
asymptotic result \eqref{eq:Meffs-MCS} of the MCS model,
apart from the sign of the linear term.

As will be seen later, these effective mass squares appear directly in
the transition amplitudes or, at least, determine their strength. But
they even occur in \emph{free} theories and might, for example, play
a role in Lorentz-noninvariant neutrino oscillations due to
Fermi-point splitting  \cite{K:neutrino}. In this case,
the  dispersion relation of the neutrino is essentially  the
same as the one of our scalar model, Eq.~\eqref{eq:omegascalar}.
As the effective mass squares (and their differences) increase
linearly with energy, the usual energy-dependence of the
oscillations is canceled. The momentum dependence of the
effective mass squares might, therefore, play an important role
in neutrino oscillations,
provided the usual Lorentz-invariant masses are small enough.

\section{Decay width and coordinate-frame independence}
\label{sec:decay}

\subsection{Decay width and decay parameter}

For the decay of a particle with momentum  $\vec q$ and
dispersion relation $\omega(\vq)$ into $n$ particles with
momenta $\vk_i$  and
dispersion relations $\omega_i(\vk_i)$, the width can be defined as
follows \cite{deWitSmith}:
\begin{eqnarray}
\label{eq:decay-width}
\Gamma(\vec q) &\defeq& \frac{1}{\sigma}  \frac{1}{N(\vec q)} \int
\left(\, \prod_{i=1}^n \frac{\id^3 k_i}
{(2\pi)^3\,N_i(\vec k_i)}\right) \nonumber\\[2mm]
 && \times\, (2\pi)^4 \delta^3\big(\vq - \sum_j \vk_j\big)\,
\delta\big(\omega(\vq) - \sum_l \omega_l(\vk_l)\big) \,
\abs{A(\vec q, \omega, \vec k_i, \omega_i)}^2\text,
\end{eqnarray}
with symmetry factor $\sigma$ for identical decay products and
transition amplitude $A$. In this expression,
$N$ and $N_i$ are normalization factors
which, in the Lorentz-invariant case, are given by $2\omega$ and
$2 \omega_i$ with corresponding normalization of the amplitude.

We find the following normalization factor
for the Lorentz-violating scalar model \eqref{eq:scalar}:
\begin{equation}
N=2 \left(\omega\pm  m\, \widehat\zeta^0\right)\geq0\text,
\end{equation}
where the (upper) lower sign is for the field ($\phi$) $\bar\phi$.
For the MCS model \eqref{eq:MCS} in the purely spacelike frame, the
normalization factor reduces to the usual one,
\begin{equation}\label{eq:twoomega}
N=2\omega\text,
\end{equation}
but with $\omega$ given by expression \eqref{eq:MCS-omega}.

The ``decay parameter'' $\gamma$ is now defined by
\begin{equation}\label{eq:gamma}
\gamma\defeq N(\vq) \,\Gamma(\vq)\text.
\end{equation}
In a Lorentz-invariant theory, $\gamma$ does not depend on the
initial three-mo\-men\-tum $\vq$
but only on the masses of the particles involved. The reason is that
the only Lorentz invariants available
are functions of the masses.\footnote{The use of the decay parameter
(or, rather, the decay constant)
instead of the decay width in the rest frame is only necessary if one
has massless particles; cf. Refs.~\cite{Havas,FM:decay}.}
In a Lorentz-violating theory, the decay parameter may also depend on the
momentum $\vec q$ through
contractions of $q^\mu$ with the background tensors.

In most cases, it suffices to calculate $\gamma$ in a
specific frame and to generalize to the coordinate-independent
expression. The calculation will be especially simple for two types of
coordinate frames:
\begin{enumerate}
\item
the class of frames with \emph{vanishing} three-momentum
of the decaying particle, $\vec q =\vec 0$, for the case of
timelike four-momentum $q^\mu$ and arbitrary background vector $\zeta^\mu$;
\item
the class of frames where the background vector is \emph{purely} spacelike
or timelike, for the case of a spacelike or timelike background vector $\zeta^\mu$.
\end{enumerate}
For example, if the calculation in our models is performed in a frame
where $\zeta^\mu$ is purely spacelike, $\gamma$ will be a function of
$\vec q \cdot \vec \zeta$, which has the
frame-independent form $-q^\mu \zeta_\mu\,$.

\subsection{Lifetime and quasi-restmass}
\label{sec:lifetime}

It appears reasonable to demand that, at least in first approximation,
the ``lifetime'' of an unstable particle, defined by
\begin{equation}
T(\vq) \defeq 1 / \Gamma(\vq)\text,
\end{equation}
transforms as a genuine time with respect to changes of frame (i.e., as the
zero-component of a four-vector). Then, the quantity $T$ must change as follows:

\begin{equation}\label{eq:lifetime}
T(\vq', \vec\beta)= \frac{1}{\sqrt{1-|\vec\beta|^2}} \,
 \big(1 - \vec \beta \cdot \vv(\vq)\big)\,T(\vq)\text,
\end{equation}
with $\vv$ the velocity of the particle [tentatively identified with the
group velocity $\vv_g \!\equiv \partial \omega/\partial \vq$
for energy $\omega(\vq)$ in the old frame],
$\vq$ and $\vq'$ the momenta in the old and new frame, and $\vec\beta$ the
boost velocity characterizing the motion of the new frame relative
to the old. Equation \eqref{eq:lifetime} can be checked explicitly
in the models of Section~\ref{sec:lorentz},
provided the group velocity is used for $\vv$.

As long as $\abs{\vv_g(\vq)} < 1$, there
always exists a ``rest frame'' with $\vec v_g'=\vec{0}$ and
\begin{equation}
\left. T \right|_\text{rest} \equiv
T(\vq, \vv_g(\vq) )= \sqrt{1-\abs{\vv_g(\vq)}^2} \,\; T(\vq)
\leq T(\vq)\text.
\end{equation}
One can take $T|_\text{rest}$, the minimum of the lifetimes over all frames,
as the definition of the observer-independent lifetime of that specific
mode. This lifetime will then be properly time-dilated
in frames where the decaying particle is moving.

For Lorentz-invariant theories, the  rest-frame decay width can be
written as
\begin{equation}\label{eq:GammaAtRest}
\Gamma_\text{rest} \equiv \left(\left. T \right|_\text{rest}\right)^{-1}
= \frac{1}{2 M_\text{rest}} \, \gamma\text,
\end{equation}
where $M_\text{rest}$ is simply the mass of the decaying particle.
By analogy, we define the mode-dependent ``quasi-restmass''
\begin{equation}
M_\text{rest} \equiv
\frac{1}{2} N(\vq) \, \sqrt {1 - \abs{\vv_g(\vq)}^2}\text,
\end{equation}
so that \eqref{eq:GammaAtRest} also holds for Lorentz-violating
theories. Note that there is no direct relation between this mass and the
effective mass square as discussed in Section~\ref{sec:lorentz}.

For the Maxwell--Chern--Simons model \eqref{eq:MCS},
we obtain
\begin{equation}\label{eq:MCS-qrm}
M_\text{rest}^{\text{(MCS)}} = \frac{m}{2} \left(1 \pm \frac{m^2}
{\sqrt{m^4 +4\, (q^\mu\zeta_\mu)^2}}\right)\text,
\end{equation}
which is always less or equal than $m$. With $\abs{q^\mu \zeta_\mu}$
increasing from zero to infinity, $M_\text{rest}^{\text{(MCS)}}$ interpolates
monotonically between $0$ and $m/2$ for the $\minus$ mode
and between $m$ and $m/2$ for the $\plus$ mode.
For the scalar model (\ref{eq:scalar}), the quasi-restmass is constant,
\begin{equation}\label{eq:scalar-qrm}
M_\text{rest}^\text{(scalar)} = \sqrt{\Mphi^2 + \zeta^2}\text.
\end{equation}

In the remainder of this article, we will focus on calculating the decay
parameter $\gamma$ as defined by Eq.~\eqref{eq:gamma},
which, as explained above,
may become momentum-dependent in Lorentz-noninvariant theories.

\section{Lorentz-noninvariant decay processes of scalars}
\label{sec:scalar}

\subsection{Preliminaries}

In this section, we show how the effective mass squares of
Section~\ref{sec:masses} enter in a simple setting. The model used is the
scalar model defined by the action \eqref{eq:scalar}.

We, first, consider decay processes with two particles in the
final state. In a Lorentz-invariant theory,
the phase-space integral would then be trivial because the
transition amplitude square
is constant. Also, we start with scalar particles in order to avoid the
complications of fields with spin.
[For particles with spin, different modes are split and
polarization identities, such as Eq.~\eqref{eq:pol-sum}
in the first appendix, may introduce the Lorentz-breaking tensors
explicitly into the amplitude; see, e.g., Eq.~\eqref{eq:prob} below.]

In addition to the Lorentz-violating particle $\phi$  with mass $\Mphi$,
we introduce a real Lorentz-invariant scalar particle $\psi$ with mass $\Mpsi$.
Its free action is given by
\begin{equation}
\action_{\psi^2}=\int_{\mathbb{R}^4} \id^4 x \,
 \Big( \frac{1}{2} \, \partial_\mu \psi\, \partial^\mu
\psi - \frac{1}{2} \,\Mpsi^2 \, \psi^2 \Big)\text.
\end{equation}
Purely for illustrative purposes, we will employ \emph{ad hoc}
Lorentz-invariant interactions between these two particles,
which can, for example, be obtained by integrating out another heavy
Lorentz-invariant particle.
The corresponding coupling constants (generically denoted $g$)
are assumed to be sufficiently small, so that the tree-level results
can be used in a first approximation.

Regarding Lorentz-noninvariance effects in tree-level decays,
there are, in general, three cases to consider:
\begin{enumerate}
\item only the decay particle violates Lorentz invariance;
\item one or more of the decay products violate Lorentz invariance,
      but not the decay particle;
\item both the decay particle and at least one of the decay products violate Lorentz
      invariance.
\end{enumerate}
We will discuss these three cases in turn.

There is also the possibility of having explicit Lorentz violation in
the interaction Hamiltonian, but this is of minor interest here,
as we intend to study how modified dispersion relations control physical
interaction processes which could naively be expected to be Lorentz invariant.

\subsection{Two-particle decay of $\phi$ into $\psi$'s}

In a general Lorentz-invariant scalar theory, the two-particle
decay $\phi\to\psi\psi$ is kinematically allowed if
the mass of the decaying particle is larger
than the sum of the masses of the produced particles.
But, for the Lorentz-violating model considered, it is really
the effective mass which determines the threshold.
With $\Meff(\qp)=\sqrt{2 m \qp + \Mphi^2} >2\Mpsi$, the following inequality
must hold in the purely spacelike frame:
\begin{equation}\label{eq:condition}
 m \qp > 2\Mpsi^2-\Mphi^2 /2 \,\text.
\end{equation}

Because only the decaying particle has Lorentz violation, the
squared transition amplitude $|A|^2$
remains constant over the two-particle phase-space, as in
the usual case. Hence, the decay parameter is simply the product of
$|A|^2$  and the two-particle phase-space volume,
\begin{equation}
V= \frac{\lambda(\Meffs, \Mpsi^2, \Mpsi^2)}{2!\;8 \pi \Meffs}\text,
\end{equation}
with the K\"allen function
$\lambda(x,y,z)\equiv \left(x^2 + y^2 + z^2 - 2xy -2yz -2zx\right)^{1/2}\,$;
cf. Ref.~\cite{deWitSmith}.
The phase-space volume $V$ depends only weakly on $\Meffs$ if
the mass $\Mpsi$ of the decay products  is small. All in all,
the decay width in the purely spacelike frame is given by
\begin{equation}
\Gamma=\frac{1}{2 \omega} \,\gamma,\quad  \gamma =  V \, |A|^2\text.
\end{equation}

One observes that the decay width behaves  as if
the decaying particle had a Lorentz-invariant mass square equal to
the effective mass square $\Meffs$ as defined  by
Eq.~\eqref{eq:Meffs-scalar}. [The decay width has a factor
$1/(2\omega+ 2\zeta^0)$
instead of $1/(2\omega)$ for an arbitrary spacelike frame.]
But, using Eqs.~\eqref{eq:GammaAtRest} and \eqref{eq:scalar-qrm},
the decay width \emph{at rest} is found to be given by
\begin{equation}
\Gamma_\text{rest}=\frac{1}{2 \, \sqrt{\Mphi^2-m^2}}  \,V\, |A|^2\text,
\end{equation}
which is larger by a factor $\Meff/\sqrt{\Mphi^2-m^2}$ than the width
of a  Lorentz-invariant particle with mass square equal to $\Meffs$.
The reason is simply that the particle $\phi$
moves with a  velocity that is different from the one of
a Lorentz-invariant particle
with the same momentum $\vec q$ and energy $\omega$ for a mass-square value
equal to $\Meffs(\vec q)$.

Next, consider special interactions of the type
\begin{equation}
\action_{\phi\psi^2}=\frac{g}{2!} \int_{\mathbb{R}^4} \id^4 x \;
\left( (\Box^\boxpower \phi) \psi^2 + \text{H.c.}\right)\text,
\end{equation}
for an integer $\boxpower \geq 1$  and a real coupling constant $g$ of mass
dimension $1-2\boxpower$. Provided condition \eqref{eq:condition} holds,
we then have
\begin{equation}
\gamma(\vq)=g^2\,V \,\Meff^{4\boxpower}(\vq)\text.
\label{gammaphipsipsi}
\end{equation}
For these special interactions, the decay width depends strongly on $\Meffs$
and, thus, on the momentum component in direction  $\widehat\vz$. This
dependence is, however, accompanied by the same power of the small
Lorentz-violating parameter $m$, according to Eq.~(\ref{eq:Meffs-scalar}).

The decay parameter \eqref{gammaphipsipsi}, given in terms of the effective
mass square \eqref{eq:Meffs-scalar}, implies a different lifetime
of the scalar and antiscalar particle (at least, at tree level),
which is consistent with \DS{CPT} noninvariance \cite{Sakurai}.
Lorentz  noninvariance is manifest from the nontrivial momentum
dependence of $\gamma(\vq)$.

\subsection{Two-particle decay of  $\psi$ into $\phi$'s}

We, now, introduce another special interaction term,
\begin{equation}\label{eq:psiTophiphi}
\action_{\psi\phi^2}=\frac{g}{2!}\int_{\mathbb{R}^4} \id^4 x \,
\left( \psi (\Box^\boxpower \phi) \phi + \text{H.c.}\right)\text,
\end{equation}
between the standard scalar particle $\psi$ and two
Lorentz-violating particles $\phi$. Interaction terms with
d'Alembertian operator $\Box$ on $\psi$ would just produce
factors $\Mpsi^2$ by use of the field equation for $\psi$.
Also, we do not consider other types of second derivatives on the two
$\phi$ fields.

The decay $\psi\to\phi\phi$ is only possible if the three-momentum
$\vq$ of $\psi$ satisfies
\begin{equation}
\qp \leq \frac{\Mpsi^2-4 \Mphi^2}{4 m}\text.
\end{equation}
The phase-space integration is difficult in a general frame.
We can, however, evaluate the phase-space integral in a
purely spacelike frame where the integral reduces to the standard one.
In this frame, the invariance with respect to boosts in
a direction orthogonal to $\vz$ allows us to focus on the
case of decay momentum $\vq\parallel\vz$, because $\gamma$
cannot depend on $\vqo$. Then, we can reduce the phase-space integral
to a one-dimensional integral over the momentum component $\kp$ of one
of the decay products,
\begin{equation}
\gamma(\vq)=\frac{1}{2!\, 8\pi \, \sqrt{\qp^2+\Mpsi^2}}\;\int\id \kp \;
\abs{A}^2\, \Big|_{\ko=\koa{0}}\text,
\end{equation}
where the integration domain is determined by energy conservation
and $\koa{0}$ is a function of the parallel components, whose
explicit form
is not needed if the tensor structure of the integrand is known.

For the scalar model considered, the amplitude is still a function of
the effective masses. But these effective masses
are not constant over the
available phase-space. The resulting decay parameter
will, in general, not be a constant, even though the decaying particle has a
Lorentz-invariant dispersion relation.

For the interaction \eqref{eq:psiTophiphi}, we find, in fact, the following
decay parameter:
\begin{equation}
\gamma(\vq)\sim  g^2 \, m^{2\boxpower} \, \qp^{2\boxpower}\text,
\end{equation}
which is consistent with the naive assumption of counting the number
of derivatives in the interaction.

\subsection{Splitting of one $\phi$ into two or more $\phi$'s}
\label{sec:splitting-phi}

The $\phi$ particle is also unstable against self-splitting, provided
\begin{equation}
\qp \leq - \frac{3 \Mphi^2}{2 m}\,\text.
\end{equation}
In this case, the Lorentz noninvariance of the decay parameter
has two origins: first, the decaying particle at the initial three-momentum
$\vec q$ and, second, the decay products averaged over the allowed
phase space.

With the special (charge-nonconserving) interaction
\begin{equation}
\action_{\phi \bar\phi^2+\bar\phi\phi^2}=\frac{g}{2!}\int_{\mathbb{R}^4}
\id^4 x\,
\left( \bar\phi^2 \, \Box^\boxpower \phi - \phi\, \bar\phi \, \Box^\boxpower
\bar \phi \; +\; \text{H.c.}\right)\text,
\end{equation}
the transition amplitude square for double-splitting becomes
\begin{equation}\label{eq:scalar-split}
\abs{A}^2\sim g^2 \left(\Meffs(\vq)^{\boxpower}+(-\Meffs(\vk))^\boxpower
+(-\Meffs(\vq-\vk))^\boxpower\right)^2\text.
\end{equation}
For $\boxpower\geq 2$,  one obtains after integration
$\gamma\sim g^2 \, m^{2\boxpower} \, \qp^{2\boxpower}$,
just as expected from counting the powers of $\qp$ and $\kp$ in
\eqref{eq:scalar-split}. For $\boxpower=1$, the leading term
in the amplitude cancels out and $\gamma$ is a constant.

\subsection{Summary}

In this section, we have seen that effective
mass squares, introduced in Section~\ref{sec:masses},
determine how the fundamental Lorentz-symmetry
breaking feeds into physical interaction processes.
In many cases, one can also see, from the derivative structure of the
interaction Lagrangian,
which powers of the Lorentz-violating parameters
appear in the transition amplitude and decay width.

\section{Vacuum Cherenkov radiation in modified QED}
\label{sec:cherenkov}

After the toy models of the previous section,
we turn to more interesting decay processes involving
Maxwell--Chern--Simons (MCS) photons.
There are, of course, no two-particle decays
in the free MCS theory and we must add charged particles
(at least, if we wish to keep the theory renormalizable).

We, therefore, enlarge the Maxwell--Chern--Simons model
by adding conventional electrons. The relevant
action is then given by
\begin{equation}\label{eq:mod-QED}
\action_\text{\,QED+spacelike\;CS--term}=
\action_\text{\,QED}+
\action_{\text{CS},\;\widehat{\zeta}^\mu\widehat{\zeta}_\mu=-1}\; \text,
\end{equation}
with the Chern--Simons-like term \eqref{eq:CSterm} for spacelike
$\widehat{\zeta}^\mu$ added to
the standard quantum electrodynamics (QED) action \cite{JauchRohrlich},
\begin{equation}\label{eq:QED}
\action_\text{\,QED}=
\action_\text{M} +\action_\text{D}\, \text,
\end{equation}
consisting of the Maxwell term \eqref{eq:Mterm} and the Dirac term
\begin{equation}\label{eq:Dterm}
\action_\text{D}= \int_{\mathbb{R}^4} \id^4 x \;
\bar \psi \,(\ii\,\gamma^\mu\partial_\mu - M - e \,\gamma^\mu A_\mu)\,\psi\text,
\end{equation}
where the electron from field $\psi(x)$ has charge $e$ and mass $M>m/2>0$.
As before, the two polarization modes of the MCS photon are denoted
$\plus/\minus$, corresponding to the $+/-$ sign in the dispersion
relation \eqref{eq:MCS-omega}.

The kinematics and interactions allow for pair creation, $\plus\to
e^+e^-$, which has been studied previously in Ref.~\cite{OS}.
But, this process has a threshold:
pair creation is only possible for large enough photon momentum in the
purely spacelike preferred frame,
\begin{equation}\label{eq:thresholdpaircreation}
\qpabs \geq \frac{2M\sqrt{4M^2-m^2}}{m} \sim \frac{4M^2}{m}\text,
\end{equation}
which corresponds to having an effective
photon mass larger than twice the electron mass.
This photon momentum is, however, far beyond the Planck scale,
\begin{equation}\label{eq:thresholdpaircreation-num}
\qpabs\gtrsim 10^{45}\unit{eV}
\left(\frac{M}{511\unit{k\,eV}}\right)^2
\left(\frac{10^{-33}\unit{eV}}{m} \right)\text,
\end{equation}
for electron mass $M$ and a ``realistic'' value
for Chern--Simons scale $m$ \cite{CFJ:MCS,CK:Lorentz}.

Photon decay into neutrinos would be an interesting alternative,
if at least one neutrino mass state is strictly massless or has a mass
$M$ of at most $10^{-7}\unit{eV}$ (corresponding, for $m=10^{-33}\unit{eV}$,
to a threshold of some $4\times 10^{19}\unit{eV}$,
close to the highest known energy of
cosmic rays). But this decay amplitude only arises at one loop,
as the neutrino has no electric charge.

\begin{figure}[t]
\begin{center}
\epsfig{file=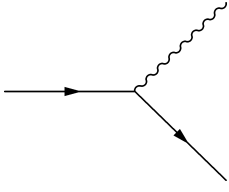, width=3cm}
\vspace*{.25cm}
 \caption{Feynman diagram contributing to vacuum Cherenkov radiation  (see text).}
\label{FDcherenkov}
\end{center}
\vspace*{.5cm}
\end{figure}

The Cherenkov process $e^-\to\minus\, e^-$, on the other hand, occurs
already at tree level and is allowed for \emph{any} three-momentum
$\vq$ of the  electron, provided $\vq \cdot \widehat\vz\ne 0$.
(See, e.g.,  Refs.~\cite{CG:cosmic,JLM:constraints} for a
general discussion of vacuum Cherenkov radiation
and Ref.~\cite{LP:Cherenkov} for a discussion in the
context of the MCS model.) The tree-level amplitude $A$ for this
process follows directly from the QED interaction,
\begin{equation}
A = \bar u(q-k)\, \bar\epsilon_\mu(k)\, (- e \gamma^\mu)\, u(q)\text,
\label{eq:cherenkov-amplitude}
\end{equation}
with $u$ the incoming and $\bar u$ the outgoing spinor and
$\bar\epsilon_\mu$ the conjugate polarization vector of the MCS photon.
The corresponding Feynman diagram is shown in Fig.~\ref{FDcherenkov}
(the Feynman rules of standard QED are given in, e.g.,
Refs.~\cite{JauchRohrlich,deWitSmith}).
Remark that the process $e^-\to\plus \, e^-$ is not allowed kinematically.
Note, furthermore, that we restrict ourselves to tree-level calculations in
the theory \eqref{eq:mod-QED} with Lorentz violation in the photon sector,
but loops involving MCS photons will probably induce Lorentz violation
also in the fermion sector.

After some $\gamma$-matrix algebra and averaging (summing) over
initial (final) spinor polarizations, one obtains
\begin{equation}
\frac12 \sum_\text{spins} \abs{A}^2 = e^2\,
\left(4 q^\mu q^\nu - 2 q^\mu k^\nu - 2 q^\nu
k^\mu + 2 q^\beta k_\beta \, \eta^{\mu\nu}\right) \,
\bar \epsilon_\mu \epsilon_\nu\text.
\end{equation}
Inserting the polarization identity \eqref{eq:pol-sum} for
$\bar\epsilon_\mu \epsilon_\nu$ and using
the on-shell relation $2 q^\beta k_\beta= k^2$
together with the photon dispersion law \eqref{eq:dispersion},
this sum simplifies to
\begin{equation}\label{eq:prob}
\frac12 \sum_\text{spins} \abs{A}^2 = \frac{e^2}{2 k^2+\zeta^2}\,
 \big( k^2 ( \zeta^2 - 4 M^2) - 2(k\cdot \zeta)^2
+ 4 (q\cdot \zeta)(k \cdot \zeta) - 4 (q\cdot \zeta)^2\big)\text.
 \end{equation}
Here, a new feature is seen to  enter two-particle decay:
the Lorentz-violation parameter
$\zeta^\mu$ is introduced explicitly into the amplitude
by the modified photon polarization identity.
Equation \eqref{eq:prob} contains moreover the contraction $q^\mu \zeta_\mu$,
although the electron itself has no direct  Lorentz violation. Hence, it is
not true that the transition amplitude square
for a two-particle decay process always reduces to a function of the
effective mass squares.

In the coordinate frame with $\zeta^0 =0$ and
for initial electron momentum $\vq$  parallel to $\vz$,
the phase-space integral can be reduced in the same way as for the scalar
models of the previous section. One finds
\begin{equation}\label{eq:gamma-cherenkov}
\gamma(\qp)=
\frac{1}{8\pi\,\sqrt{\qp^2+M^2}}\;\int_0^{k_\text{max}} \id \kp \;\,
\frac12\sum_\text{spins} |A|^2 \,\Big|_{\ko=\koa{0}},
\end{equation}
where $\koa{0}$ is a known function of $\qp$ and $\kp$ and the integration of
the photon momentum component $\kp$ runs
from zero to a (positive or negative) value $k_{\text{max}}$ defined by
\begin{equation}\label{eq:kmax}
k_{\text{max}}(\qp) \equiv \frac {2 m  \qp\,\left(m +2\, \sqrt{\qp^2+M^2}\right)}
{m^2+4M^2+4 m\, \sqrt{\qp^2+M^2}}\,\text.
\end{equation}
In the coordinate frame with $\zeta^0 =0$ and $|\vz|=m$,
the decay width is then
\begin{equation}
\Gamma(\vq)=\frac{1}{2\,\sqrt{|\vq|^2+M^2}}\; \gamma(\qp)\text.
\end{equation}
Equation \eqref{eq:gamma-cherenkov} can be integrated analytically and
the decay parameter becomes
\mathindent=0mm
\begin{equation}\begin{split}\label{gammacherenkov}
\gamma(\qp)=&\frac{\alpha}{16\,\sqrt{\qp^2+M^2}}\,
 \Bigg[
 2 m |k_\text{max}|\, \sqrt{m^2 + 4 k_\text{max}^2}
 - 8 m |\qp| \left(\sqrt{m^2+4k^2_\text{max}}-m\right)
 \\[2mm]
 &   -4  \left(m^2 + 4M^2\right) |k_\text{max}|
  + m \left(m^2 + 8 M^2 + 16 \qp^2\right)
  \arcsinh\left(\frac{2 |k_\text{max}|}{m}\right)
  \Bigg]\,\text,
\end{split}\end{equation}
\mathindent=2em  
with  fine-structure constant  $\alpha\defeq e^2/(4\pi)$
and $k_\text{max}$ from Eq.~\eqref{eq:kmax}.

For $0\leq\qpabs<M$, the result \eqref{gammacherenkov} can be expanded
in $m/M$,
\begin{equation}\label{gammacherenkov-lowE}
\gamma(\qp)=
(4/3)\,\alpha\; m\,\qpabs^3/ M^2
 + \BigO\left( \alpha\, m^2\, \qpabs^3 / |M|^3 \,\right)\text,
\end{equation}
while, for $\qpabs\gg M$, an expansion in $m/\qpabs$ and $M/\qpabs$ gives
\begin{equation}\label{gammacherenkov-highE}
\gamma(\qp)=
\alpha \,m \, \qpabs \, \big(\ln(\qpabs/m)+2\ln 2-3/4\big)+ \cdots\,\text,
\end{equation}
where the ellipsis stands for subdominant terms.
Hence, the decay parameter of the electron grows approximately linearly
with the momentum component in the preferred direction, but is
suppressed by one power of $m$, unlike the result in other theories
where the decay rate increases more strongly above threshold
\cite{CG:cosmic,JLM:constraints}.

For a relativistic electron, $|\vec q|\gg M$, the Cherenkov amplitude
\eqref{eq:cherenkov-amplitude} preserves the helicity of the electron.
While the complete decay parameter $\gamma$ turns out to be different for
left- and right-handed electrons, the leading term proportional to
$\alpha \,m \,\qpabs \, \ln(\qpabs/m)$ is equal for both helicities.

The emitted $\minus$
photon is approximately left- or right-circularly polarized
(depending on the sign of $\kp$) for momentum component $\kpabs\gg m$, while,
for  $\kpabs/m\to 0$, its polarization becomes linear \cite{K:CPT-review2005}.
Combined with electron helicity conservation, this implies that the
Cherenkov decay violates angular momentum conservation by approximately one
unit for large photon momentum component $\kpabs$.
This is reflected in the $|\qp| \gg M$ transition amplitude
which is strongly peaked at $\kpabs \lesssim m$.

The vacuum Cherenkov decay process has also been studied
quantum mechanically in Ref.~\cite{LP:Cherenkov} for
modified QED \eqref{eq:mod-QED} with a ``lightlike''
Chern--Simons-like term.\footnote{For the purely spacelike case,
the authors of Ref.~\cite{LP:Cherenkov} have furthermore calculated
the classical radiation rate, effectively in the limit $M\to\infty$.
They obtain a rate  proportional to  $Q^2\,|\vz_\text{class}|^2$,
where $Q$ is the classical charge  and $\vz_\text{class}$
the classical Chern--Simons parameter with dimension of inverse length.
Our calculation reproduces their result (21) with corrections of order
$Q^2 \, |\vz_\text{class}|^2\, m/M$, for $m =\hbar\, |\vz_\text{class}|/c$.}
An estimate for the decay width of an electron
at rest ($\zeta^\mu$ is lightlike) has been obtained \cite{LP:Cherenkov}
by effectively expanding in $\abs{\,\zeta^0/M}=\abs{\,q^\mu\zeta_\mu} / q^2$,
\begin{equation}\label{eq:LP-result}
\Gamma_\text{rest}^\text{\,(QED+lightlike\;CS--term)}
\overset{?}\sim
\,\kappa \, \frac{\alpha}{2 M}\, \abs{\,q^\mu \zeta_\mu}\text,
\end{equation}
with an unknown constant $\kappa$.
This behavior agrees with the expansion in $\abs{\,q^\mu
\zeta_\mu} / q^2$ of the exact result \eqref{gammacherenkov}.
In the rest frame of the electron, our result gives, namely,
\begin{equation}\label{eq:KK-result}
\Gamma_\text{rest}^\text{\,(QED+spacelike\;CS--term)}
\sim \frac{\alpha}{2 M}\,  \abs{\,q^\mu \zeta_\mu} \,
\big(\ln (\abs{\,q^\mu \zeta_\mu} / m^2) + 2\ln 2 - 3/4\big)\text.
\end{equation}
Equation \eqref{eq:KK-result} contains an additional logarithmic
dependence on $\abs{\,q^\mu \zeta_\mu}$, which may, however, be absent
for the lightlike MCS model because a lightlike vec\-tor $\zeta_\mu$
does not define a mass scale $m$.

\section{Photon triple-splitting in modified QED}
\label{sec:photon}

At last, we turn to photon triple-splitting from the purely spacelike
MCS model \eqref{eq:MCS}, continuing the work of
Ref.~\cite{AK:photon-decay}. There are eight decay channels, corresponding
to all possible combinations of the different modes $\plus$ and
$\minus$ from  dispersion relation \eqref{eq:MCS-omega}.
In Appendix~\ref{ap:kinematics}, we show that the following three
cases are allowed for generic initial three-momentum~$\vq$:
$\plus\to\minus\minus\minus$, $\plus\to\plus\minus\minus$, and
$\minus\to\minus\minus\minus$, whereas the five others are
kinematically forbidden. For special momentum $\vq \perp \vz$,
the decay rate turns
out to be zero except for the case of $\plus\to\minus\minus\minus$.

The implication would be that, with suitable interactions,
all MCS photons are  generally unstable against splitting.
The exception would be for the lower-dimensional subset
of  $\minus$ modes with three-momenta orthogonal to $\vec\zeta$.

\begin{figure}[t]
\begin{center}
\epsfig{file=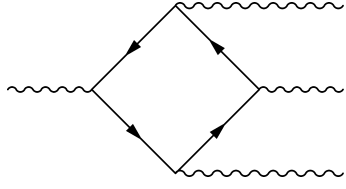, width=4.5cm}
\vspace*{.25cm}
 \caption{Feynman diagram contributing to photon triple-splitting (see text).}
\label{FDphotontriplesplitting}
\end{center}
\vspace*{.5cm}
\end{figure}

Following Ref.~\cite{AK:photon-decay}, the interaction is taken to be the
Euler--Heisenberg interaction and the photonic action considered reads
\begin{align}\label{eq:MCSEHaction}
\action_\text{photon}&=
\action_{\text{MCS},\;\widehat{\zeta}^\mu\widehat{\zeta}_\mu=-1\;,\;\widehat{\zeta}^0=0}
+\action_\text{EH}\text,
\end{align}
consisting of the MCS quadratic term \eqref{eq:MCS},
for purely spacelike background four-vector $\widehat{\zeta}^\mu$,
and the quartic Euler--Heisenberg term
\begin{align}\label{eq:EHterm}
\action_\text{EH}&=K \int_{\mathbb{R}^4} \id^4 x\, \left[\left(
\,\frac12 \, F_{\mu\nu}F^{\mu\nu}\right)^2
+7\left(\,\frac18 \, \epsilon_{\mu\nu\rho\sigma}
F^{\mu\nu}F^{\rho\sigma}\right)^2\right]\text,
\end{align}
with effective coupling constant
\begin{equation}\label{eq:Kdef}
K \equiv \frac{2\alpha^2}{45 M^4}\text,
\end{equation}
given in terms of the fine-structure constant
$\alpha\equiv e^2/(4\pi)\approx 1/137$ and the electron mass $M$.
In the modified version of quantum electrodynamics (QED)
with action \eqref{eq:mod-QED},
the Euler--Heisenberg term arises from  the low-energy limit
of the  one-loop electron contribution (cf. Fig.~\ref{FDphotontriplesplitting})
to the effective gauge field action  \cite{JauchRohrlich}.

As we work in a purely spacelike frame, the
usual definition of the phase-space integral is applicable,
according to the discussion of Section~\ref{sec:decay}. We only have
to perform the standard three-particle phase-space integral.
The details of the calculation are relegated to
Appendix~\ref{ap:photon}.

In the purely spacelike frame ($\zeta^0 =0$ and $|\vz|=m$),
the decay width is found to be given by
\begin{equation}\label{eq:decay-width-photon-splitting}
\Gamma(\vq)= \frac{1}{2 \omega(\vq)}\;\gamma(\qp)\text,
\end{equation}
with the following behavior of the decay parameter for $\qpabs \gg m$:
\begin{equation}\label{eq:gamma-photon-splitting}
\gamma(\qp)\sim c \, K^2 \, m^5 \qpabs^5
= c \,(2/45)^2\; \alpha^4 \; m^5 \qpabs^5/M^8 \text,
\end{equation}
in terms of the model constant $K$ from \eqref{eq:Kdef} and
a numerical constant $c$ depending on the decay channel,
\begin{subequations}\label{eq:c-numresults}\begin{align}
c(\plus\to\minus\minus\minus)  &\approx 1.078\times 10^{-7}\text,
\label{constPtoMMM}\\[2mm]
c(\minus\to\minus\minus\minus) &\approx 1.182 \times 10^{-7}\text,
\label{constMtoMMM}\\[2mm]
c(\plus\to\plus\minus\minus)   &\approx 6.214 \times 10^{-8}\text.
\label{constPtoPMM}
\end{align} \end{subequations}

The appearance of a fifth power of the momentum component $\qpabs$
in the asymptotic result \eqref{eq:gamma-photon-splitting}
can be understood as follows. First, the phase-space volume grows
linearly with $\qpabs$. Second, the four derivatives in the
interaction lead to an amplitude square
that is eighth-order in the momenta and
the phase-space integral gives a quartic dependence on the effective mass square,
which results in a fourth power of $\qp$
(see Section~\ref{sec:splitting-phi} for similar results in the scalar model).

It is, however, surprising that there are no
suppression effects from the
polarizations, since these decays violate angular-momentum conservation.
In the decay process $\plus\to\minus\minus\minus$ with initial
momentum $\vq \parallel \vz$, for example, a left-circularly
polarized photon turns into three roughly right-circularly polarized
photons with  the same total energy-momentum.

Cancellations may, however, occur for an interaction with more derivatives.
Replacing $\action_\text{EH}$ in \eqref{eq:MCSEHaction} by
\begin{equation}
\action^\prime_{A^4}=K'\, \int_{\mathbb{R}^4} \id^4 x\;
 F_{\mu\nu}F^{\mu\nu} F_{\rho\sigma} \Box F^{\rho\sigma}\text,
\end{equation}
for example, gives the following additional factors $f$ in the
transition amplitude square:
\begin{subequations}
\begin{align}
f(\plus\to\minus\minus\minus) &=
m^2 \big(\omegap(q)+\sum_i\omegapa{i}\big)^2
\sim 2 m^2 \qp^2\text, \\
f(\minus\to\minus\minus\minus) &=
m^2 \big(-\omegap(q)+\sum_i\omegapa{i}\big)^2\sim m^4\text, \\
f(\plus\to\plus\minus\minus) &=
m ^2\big(\omegap(q)+\omegapa{1}+\omegapa{2}-\omegapa{3}\big)^2\text,
\end{align}
\end{subequations}
with the asymptotic behavior for $\qpabs \gg m$ shown in two cases.
The decay channel  $\plus\to\minus\minus\minus$ then picks up
an additional factor $2 m^2 \qp^2$ compared to the
Euler--Heisenberg result,
so that $\gamma \sim c'\,  K'^2\,  m^7\, \qpabs^7$.
The decay channel $\minus\to\minus\minus\minus$, on the other hand, has
$\gamma \sim c' \, K'^2 \, m^9 \, \qpabs^5$,
with two powers of $\qp$ suppressed.

Let us return to the decay parameter $\gamma$
from the original low-energy effective theory \eqref{eq:MCSEHaction}.
The validity domain of the effective theory can be expected to be
given by $m \qpabs  \ll 4 M^2$, for
the particular Lorentz-violating decay process considered.\footnote{The
range quoted is suggested by
two results. First, we have estimated the leading-order correction to
\eqref{eq:gamma-photon-splitting}
from Feynman box diagrams (Fig.~\ref{FDphotontriplesplitting})
and find a relative correction of order $m \qpabs/M^2$.
Second, the pair-creation threshold \eqref{eq:thresholdpaircreation}
sets a boundary beyond which the Euler--Heisenberg effective
action is certainly no longer valid;
see the discussion in, e.g., Section~13.2 of Ref.~\cite{JauchRohrlich}.
Note that a similar conclusion on the validity domain
of the Euler--Heisenberg effective action
in the context of other Lorentz-violating theories
appears to have been reached in  Ref.~\cite{GNY:photon-splitting}.}
If correct, this would imply that the decay parameter
\eqref{eq:gamma-photon-splitting} grows to a value of order $\alpha^4\,M^2$.
But, it is also clear that, as long as $m \ll M$, this would require very
large energies on the scale of usual elementary particle physics;
see, in particular, Eq.~\eqref{eq:thresholdpaircreation-num} of the
previous section. Still, there may be entirely different applications of
the simple model \eqref{eq:MCSEHaction},  which may turn out to be
experimentally accessible, either directly or indirectly. As mentioned
before, the intention of the present article is purely theoretical.

With that intention stated, let us comment on the possible high-energy behavior
of photon triple-splitting in modified QED \eqref{eq:mod-QED}, as our calculation
was only valid for momenta less than the electron mass $M$ or
perhaps even for momenta less than $M^2/m$, with an extra factor $M/m$ for
Chern--Simons scale $m$. Recall that, for  standard QED,
the amplitude of a four-photon interaction to order $\BigO(\alpha^2)$
is known in principle \cite{JauchRohrlich}.

For the MCS photon, the
triple-splitting phase-space volume grows as $m \qpabs$ for $\qpabs\to\infty$.
But, for a photon-splitting process, it is also known that
all scalar products between the photon momenta vanish
in standard QED and that the amplitude
square is zero \cite{FM:decay}. Therefore, the dimensionless
MCS decay amplitude square must
be of order $m^n$, for $n \geq 1$,
as it is in the low-energy region. However, the precise
functional dependence may be rather subtle. There could, for example, be
a factor $\big(m\qpabs /(m\qpabs +M^2)\big)^4$ in the amplitude square.
For the moment, we just make the following conjecture:
\begin{equation}\label{eq:ampl-conjecture}
|A|^2\;\Big|_{\;\qpabs \gg 4M^2/m}
\stackrel{?}{\sim}\; \alpha^4\, \,\text,
\end{equation}
neglecting a possible logarithmic dependence on $m\qpabs/M^2$.

The conjectured behavior \eqref{eq:ampl-conjecture} would imply for the
decay parameter:
\begin{equation}\label{eq:gamma-conjecture}
\gamma\;\Big|_{\;\qpabs \gg 4M^2/m}
\stackrel{?}{\sim}\; c_\infty\,\alpha^4\,m\,\qpabs  \,\text.
\end{equation}
Combined with the low-energy result \eqref{eq:gamma-photon-splitting},
this would mean that the effect of Lorentz breaking
keeps growing with energy. At ultra-high energies,
the decay rate \eqref{eq:decay-width-photon-splitting}
would then approach a direction-dependent constant (up to logarithms).
A similar behavior has been seen for
vacuum Cherenkov radiation; cf. Eq.~\eqref{gammacherenkov-highE}.

\section{Discussion}
\label{sec:discussion}

In this article, we have considered photon and scalar models with a
super-renormalizable term containing a Lorentz-violating background
four-vector $\zeta^\mu \equiv m \, \widehat\zeta^{\,\mu}$.
The modified dispersion relations of these models shift the energy only
slightly for high momentum $\abs{\vq}\gg m$. In the spacelike
Maxwell--Chern--Simons model \eqref{eq:MCS}, for example, one has
\begin{equation}\label{mildLV}
\omega(\vq)\sim \abs{\vq} \pm \, \abs{\cos\theta}\, m/2\text,
\end{equation}
where $\theta$ is the angle between wave vector $\vq$ and the background
vector $\vz \equiv m \, \widehat\vz$ in a purely spacelike frame.
This dispersion relation is different from other
Lorentz-violating dispersion relations considered in the literature
(e.g., Refs. \cite{GNY:photon-splitting,JLM:constraints})
where the deviation from Lorentz invariance grows rapidly with momentum.

But even the mild Lorentz violation \eqref{mildLV}
has important effects on the
high-energy behavior as the absolute value of the effective
mass square \eqref{eq:Meffs-MCS} does increase with momentum. Both the
phase-space measure and the derivative terms in the interaction pick up
this effective mass square.

For theories with only scalar particles, the Lorentz-violating effects
can be understood solely in terms of the modified kinematics.  For particles
with spin, the amplitude is also modified by the different tensor structure
of the photons; see, in particular, the discussion below Eq.~\eqref{eq:prob}.

The importance of this kind of Lorentz-symmetry breaking should, however,
not be overestimated because observables will still
be suppressed by powers of the soft Lorentz-breaking scale $m$.
Moreover, in the low-energy region, the Lorentz violation can be hidden
by standard mass terms that dominate the amplitude and kinematics.

Lorentz-noninvariant effects may, on the other hand, become visible
if a standard mass term is not allowed (as by gauge invariance
in the Maxwell--Chern--Simons model) or if processes are considered that are
normally kinematically forbidden (as vacuum Cherenkov
radiation and photon triple-split\-ting in standard quantum electrodynamics).
In both cases, the background vector $\zeta^\mu$ is crucial to obtain a non-zero
probability.

Up till now, we have considered one particular modification of
quantum electrodynamics (QED),
namely the theory \eqref{eq:mod-QED}
with an additional photonic Chern--Simons-like term \eqref{eq:CSterm}.
Lorentz violation in the electron sector may also lead to a nonzero
amplitude for photon-triple splitting \cite{KP:photon}. However,
without modified photon kinematics, this still does not give
a nonzero  decay width of the photon
because the phase-space volume is zero \cite{AK:comment}.
On the other hand, if there is a bi-linear Lorentz-violating contribution
to the photon action, one typically expects both photon triple-splitting and
vacuum Cherenkov radiation to occur.

Apart from the Chern--Simons-like term, there is only one other
renormalizable bi-linear Lorentz-violating term in
the Standard Model extension \cite{CK:Lorentz,KM:photon}
\begin{equation}\label{eq:other-photon-action}
\action_k = \int_{\mathbb{R}^4} \id^4 x \;
k^{\mu\nu\rho\sigma} \,F_{\mu\nu}F_{\rho\sigma}\,\text,
\end{equation}
with a real dimensionless background tensor $k^{\mu\nu\rho\sigma}$.
This background tensor $k^{\mu\nu\rho\sigma}$ has certain obvious
symmetries (the same ones as the Riemann tensor),
is taken to be doubly traceless
(so as not to give the standard Maxwell term),
and does not contain a totally antisymmetric part
(which would produce a total derivative). But, its tensor structure remains
complicated and explicit calculations of decay processes are difficult.
In the following, we will refer to \eqref{eq:other-photon-action}
as the $k$--term.

This  \DS{CPT}--even $k$--term contains, in fact, two derivatives and
yields a dispersion relation with stronger Lorentz violation at large momenta
than the \DS{CPT}--odd Chern--Simons-like term \eqref{eq:CSterm} with
a single derivative. Purely on dimensional grounds, one can write
\begin{equation}
\omega(\vq)^2 = |\vq|^2 \, \big(1 + \Theta(\widehat \vq) \big)\,\text,
\end{equation}
with $\widehat \vq \equiv \vq / \abs{\vq}$ and a dimensionless function
$\Theta$ carrying the direction dependence.
Typically, this dispersion relation leads to a direction-dependent group
velocity, unless $\Theta$ is independent of $\widehat \vq$.

Consistency of the theory demands, most likely, a group velocity
of electromagnetic waves not larger than one (here, the maximum attainable
velocity of the electrons) and the model will probably only be
causal for certain choices of background tensor $k^{\mu\nu\rho\sigma}$.
Indeed, one such choice has been presented in
Eq.~(50) of Ref.~\cite{CK:Lorentz}. For that particular
model, there is one photon mode with lightlike  momentum and one  mode
with spacelike momentum, making both photon triple-splitting and Cherenkov
radiation possible in principle.

For QED with the additional photonic term \eqref{eq:other-photon-action},
\begin{equation}\label{eq:other-mod-QED}
\action_\text{\,QED+$k$--term}=
\action_\text{\,QED}+\action_k \, \text,
\end{equation}
one expects the triple-splitting decay width to carry a larger power of the
momentum than the decay width from the theory \eqref{eq:mod-QED}
considered in the rest of this article.
The reason is that the $k$--term \eqref{eq:other-photon-action} contains
more derivatives than the Chern--Simons-like term \eqref{eq:CSterm}
and  has, moreover, no explicit mass scale
(the photonic sector violates Lorentz invariance, but, at tree level,
still has conformal invariance).
In the low-energy region of the photon-triple-splitting  process,
we conjecture that the leading term of the decay parameter is given by
\begin{equation}\label{lowEdecayparam-other-mod-QED}
\gamma(\vq)\;\Big|_{\;\kqq \ll M^2}
\overset{?}\sim \;
\alpha^4 \,\kqq^5\,/M^8\,\text,
\end{equation}
where $M$ stands for the electron mass and
$\kqq^5$ is a function with a possibly complicated tensor
structure involving $k^{\mu\nu\rho\sigma}$ to the fifth  power
and $q^\mu$ to the tenth. For the asymptotic behavior at large momentum,
we conjecture the following momentum dependence (neglecting logarithms):
\begin{equation}\label{highEdecayparam-other-mod-QED}
\gamma(\vq)\;\Big|_{\;\kqq \gg M^2}
\overset{?}\sim\; \alpha^4 \,\kqq^5/\kqq^4
\sim \,           \alpha^4 \,\kqq \,\text,
\end{equation}
with the highly symbolic notation $\kqq$.
This conjectured decay parameter \eqref{highEdecayparam-other-mod-QED}
from the $k$--term
would be qualitatively the same as \eqref{eq:gamma-conjecture}
from the Chern--Simons-like term, with possibly interesting implications
for the high-energy theory. However, only a complete calculation can tell
whether or not \eqref{eq:gamma-conjecture}
and \eqref{highEdecayparam-other-mod-QED} hold true.

\ack

It is a pleasure to thank E. Kant and C. Rupp for useful discussions.

\begin{appendix}
\section{Photon polarization vectors in the MCS model}\label{ap:polarizations}

In this appendix, we present simplified expressions for the
polarization vectors of the Maxwell--Chern--Simons (MCS)
gauge field in purely spacelike frames.
For explicit calculations involving multiple particles, it turns out
to be more
practical to have them in a momentum-independent basis, unlike the
expressions used in Refs.~\cite{AK:photon-decay,LP:Cherenkov}.

Let $\widehat\vx$ be an arbitrary unit vector orthogonal to the
background three-vector $\widehat\vz$ and define
$\widehat\vy \defeq \widehat\vz \times \widehat\vx$.
Then, use cylinder coordinates with $\widehat\vz$ as the cylinder axis
and azimuthal angle $\phi$ measured away from $\widehat\vx$. The momentum
three-vector $\vk$ can now be written as
\begin{equation}
\vk = \kp \, \widehat\vz + \ko  (\cos\phi \; \widehat\vx
+ \sin\phi \; \widehat\vy) = R_\phi\,(\ko, 0, \kp)^\text{T}\text,
\end{equation}
with $\text{T}$ standing for `transpose' and the rotation matrix
\begin{equation}\label{eq:Rphi}
R_\phi \defeq\begin{pmatrix}
\cos \phi & -\sin \phi &\;\; 0 \\
\sin \phi & \phantom{-}\cos \phi &\;\; 0 \\
0 & 0 &\;\; 1 \\
\end{pmatrix}\text.
\end{equation}

For the gauge field $A^\mu$, the polarizations are most transparent
in the Lorentz gauge $k_\mu \epsilon^\mu(k)=0$:
\begin{equation}\label{eq:epsilon}
\epsilon^\mu_\pm(\vk)=
\frac{1}{\sqrt{ (2\, \omega_{\parallel,\pm} \mp m)\, \omega_{\parallel,\pm} }}  \;
\tilde R_\phi\left(
-\ko    \,,\,
-\omega_\pm \,,\, \pm \ii \, \omegapa{\pm} \,,\,
0\right)^\text{T}\text,
\end{equation}
which is consistent with the polarization vectors of
Ref.~\cite{LP:Cherenkov}. Here, $\tilde R_\phi$ is the rotation
matrix \eqref{eq:Rphi} extended to four-dimensional spacetime,
\begin{equation}
\tilde R_\phi \defeq
\begin{pmatrix}
1\;\;      & \vec 0^\text{T}  \\
\vec 0\;\; & R_\phi \\
\end{pmatrix}
\text,
\end{equation}
and we have used the abbreviations \eqref{eq:MCS-omega} and
\eqref{eq:MCS-omegaparallel} for the frequencies
$\omega_\pm$ and $\omegapa{\pm}$, which are taken to
have an implicit argument $\vk$.

The explicit polarizations \eqref{eq:epsilon} are not needed, as long as
the calculation produces gauge-invariant expressions of the form
\begin{equation}
\bar\epsilon^\mu(k) \epsilon^\nu(k) \, R_{\mu\nu}\,\text,
\end{equation}
with $k^\mu R_{\mu\nu}=0$ and $k^\nu R_{\mu\nu}=0$.
One can then simply replace
\begin{equation}\label{eq:pol-sum}
\bar\epsilon^\mu (k) \, \epsilon^\nu (k) \, \mapsto \,
\frac{1}{2 k^2 + \zeta^2}\big(-k^2 \eta^{\mu\nu}
- \zeta^\mu \zeta^\nu
+ \ii \, \epsilon^{\mu\nu\rho\sigma}\zeta_\rho k_\sigma\big)\text,
\end{equation}
for the $\plus$ and $\minus$  modes separately (provided $m \ne 0$).
This is analogous to the replacement
\begin{equation}
\sum_r \bar\epsilon^\mu_r(k)\,\epsilon^\nu_r(k)\, \mapsto \,
- \eta^{\mu\nu}
\end{equation}
for standard photons \cite{JauchRohrlich}.
An expression like \eqref{eq:pol-sum} also appears in the MCS photon
propagator, together with terms that vanish
on-shell (see, e.g., Ref.~\cite{AK:causality}).

Equation \eqref{eq:pol-sum} can be proven by establishing the equality
in the Lorentz gauge. Remark that, even for large $k^2$, it is not a good
approximation to drop the $\zeta^\mu \zeta^\nu$ term in
\eqref{eq:pol-sum}  because $(k^\mu \zeta_\mu)^2$ is still
approximately equal to $k^4$ according to Eq.~\eqref{eq:dispersion}.

In principle, \eqref{eq:pol-sum} suffices to
calculate every possible gauge-invariant amplitude square,
but, for completeness, we also list the electric and magnetic field
polarizations as defined in Ref.~\cite{AK:photon-decay}:
\mathindent=0.5em
\begin{equation}\label{eq:pol-f-b}
\vf_\pm(\vk)=N \, R_\phi \begin{pmatrix}
-\ii\,  \omegapa{\pm}\sqroot{\omegapa{\pm}}\,\\
\,\mp \omega_\pm \sqroot{\omegapa{\pm}}\,\\
\, \ii\, \chi\,  \ko \sqroot{\omegapa{\pm} \mp m}
\end{pmatrix},\,\;\;
\vb_\pm(\vk)= N \, R_\phi \begin{pmatrix}
\pm \chi \,\omegapa{\pm}\sqroot{\omegapa{\pm} \mp m}\,\\
\,-\ii\,\chi\, \omega_\pm\sqroot{\omegapa{\pm} \mp m}\,\\
\,\mp \ko \sqroot{\omegapa{\pm}}
\end{pmatrix}\text,
\end{equation}
\mathindent=2em
with common normalization factor $N \equiv (m^2+4\kp^2)^{-1/4}$
and $\chi \equiv \sgn \kp$.

\section{Kinematics of MCS photon triple-splitting}
\label{ap:kinematics}

In this appendix, we show how to solve the energy-momentum
conservation condition in photon triple-splitting,
\begin{equation}\label{eq:conservation}
\omega(\vq)\big|_{\,\vq=\sum_i \vk_i} =
\sum_j \omega(\vk_j)\,\text,
\end{equation}
where $\vq$ is the three-momentum of the decaying particle and
 $\vk_i\, (i=1\dots3)$ are those of the decay products. The
 energies are given by the Maxwell--Chern--Simons
 dispersion relation \eqref{eq:MCS-omega},
i.e., we work in a frame where $\widehat\zeta^{\,\mu}$ is purely spacelike.
The considerations of this appendix are independent of the kind of
interaction responsible for the splitting process.

As seen in Ref.~\cite{AK:photon-decay}, a direct solution of
\eqref{eq:conservation} is not possible.
The crucial observation, now, is that the purely spacelike theory
is still invariant under Lorentz boosts in directions
orthogonal to the preferred axis $\widehat\vz$. [The reason is that
the orthogonal momentum $\vko$ enters \eqref{eq:MCS-omega} in the
standard way.] There, then, exists a Lorentz transformation which
transforms the case $\vq \nparallel\widehat\vz$
to one with $\vq\parallel\widehat\vz$,
except for the special case of a $\minus$ photon
with $\vq \perp \widehat\vz$, which will be dealt with later.

For the case $\vq\parallel\widehat\vz$, the optimal situation obviously has
all momenta aligned because orthogonal components
of the $\vk_i$ increase the right-hand side of
\eqref{eq:conservation}. Therefore, it suffices to consider the
case with all four vectors $\vq$ and $\vk_i$
parallel to $\widehat\vz$. An elementary calculation shows then that
\begin{equation}
\omega_+(\vq) < \omega_+(\vk_1)+\omega_+(\vk_2)+\omega_-(\vk_3)\text.
\end{equation}
Hence, the decay channel $\plus\to\plus\plus\minus$ is  kinematically
forbidden and the same holds for the channel
$\minus\to\plus\minus\minus$ and those involving more $\plus$ decay
products.

On the other hand, the splitting $\minus\to\minus\minus\minus$
is allowed, because the following inequality
holds for arbitrary momentum $\vq\neq \vec{0}$:
\begin{equation}
\omega_-(\vq) > 3 \, \omega_-(\vq/3)\text.
\end{equation}
Now, $\plus\to\minus\minus\minus$ decay is also possible,
as $\omega_+>\omega_-$ holds generally. By making use of the relation
\begin{equation}
\omega_+(\vq)=\omega_-(\vq)+m \text,
\end{equation}
for $\vq\parallel\widehat\vz$, one can furthermore show that the
channel $\plus\to\plus\minus\minus$ is allowed.

There remains one special case to be discussed. If the initial
momentum of a $\minus$ photon is orthogonal to $\widehat\vz$,
there is no Lorentz transformation which removes the orthogonal
component $\vqo$. However, for $\vq \cdot \widehat\vz =0$,
the dispersion relation of the $\minus$ photon
is that of a usual massless
particle while the $\plus$ mode has mass $m$. Thus, only the splitting
$\minus\to\minus\minus\minus$ is possible here and the
allowed region has zero phase-space volume, which,
under quite general assumptions, yields a vanishing decay rate
\cite{FM:decay}. Similarly, one sees that the decay width for
$\plus\to\plus\minus\minus$ is zero for
initial three-momentum orthogonal to $\vz$.

To summarize, we have established that only three decay channels
are allowed
for photon triple-splitting: $\plus\to\plus\minus\minus$,
$\minus\to\minus\minus\minus$ and $\plus\to\minus\minus\minus$,
where the former two are for $\vq \cdot \widehat\vz \neq 0$
and the latter one for arbitrary initial momentum $\vq$. Hence, there
are no stable $\plus$ photons in the model \eqref{eq:MCS}, while the
$\minus$ photons are only stable for a momentum subset of measure
zero.

\section{Phase-space integration for MCS--EH photon triple-splitting}
\label{ap:photon}

In this appendix, we sketch the calculation of the decay parameter
$\gamma$ for photon triple-splitting in the low-energy
effective photon theory \eqref{eq:MCSEHaction}, consisting of the free
Maxwell--Chern--Simons model \eqref{eq:MCS}
with Euler--Heisenberg interaction \eqref{eq:EHterm}.

In the purely
spacelike frame, we have to perform a standard phase-space integral,
\mathindent=0.25em
\begin{align}\label{eq:photon-phase-space-integral}
\Gamma(\vec q) =&  \frac{1}{\sigma}\frac{1}{2\omega(\vec q)} \int
\left(\,\prod_{i=1}^3 \frac{\id^3 k_i}
{(2\pi)^3\, 2 \omega_i(\vec k_i)}\right) \nonumber\\[2mm]
& \times\, (2\pi)^4 \delta^3\big(\vq - \sum_j \vk_j\big)\,
\delta\big(\omega(\vq) - \sum_l \omega_l(\vk_l)\big) \,
\abs{A(\vec q, \omega, \vec k_i, \omega_i)}^2 \text.
\end{align}
\mathindent=2em
With the Euler--Heisenberg interaction \eqref{eq:EHterm},
the transition amplitude for the splitting of one photon with
momentum $\vq$ into three with momenta $\vk_i$
can be written as  \cite{AK:photon-decay}
\begin{equation}
A = -\frac{1}{2\pi^2} \, K \, (A_{12} + A_{23} + A_{31})\text,
\end{equation}
in terms of the coupling constant $K$ from  \eqref{eq:Kdef}
and the following expression:
\mathindent=0em
\begin{eqnarray}
A_{ab}&=&\Big[\big(
\vf(\vk_a)\cdot\vf(\vk_b)-\vb(\vk_a)\cdot\vb(\vk_b)\big)
\big(\vf(\vk_c)\cdot\vf^*(\vq)-\vb(\vk_c)\cdot\vb^*(\vq)\big)
\nonumber\\[2mm]
&&+\frac74\,\big(\vf(\vk_a)\cdot\vb(\vk_b)
+\vf(\vk_b)\cdot\vb(\vk_a)\big)
\big(\vf(\vk_c)\cdot\vb^*(\vq)+\vb(\vk_c)\cdot\vf^*(\vq)\big)\Big]\text,
\end{eqnarray}
\mathindent=2em
which depends on the  polarization vectors
\eqref{eq:pol-f-b} for the considered channel
(particle indices $a$, $b$, $c$ $\in\{1,2,3\}$ all different).

The evaluation of the phase-space integral for arbitrary initial
momentum is simplified by the same argument that helps with
solving the kinematics, namely, the result must be invariant with
respect to Lorentz boosts in a direction orthogonal to $\widehat\vz$.
This suggests splitting the momentum integrations as follows:
\begin{equation}
\frac{\id^3 k}{\omega(\vk)}=
\frac{\id^3 k}{\sqrt{\omegap(\kp)^2+\ko^2}}=
\id \kp \, \frac{\id^2 \ko}{\sqrt{\mu(\kp)^2+\ko^2}}\text,
\end{equation}
with $\mu(\kp)\defeq\omegap(\kp)$.\footnote{Note that this definition
has nothing to do with the effective mass square from
Section~\ref{sec:masses}.}
The momentum-conservation $\delta$-function is split in the same
way. Then, the phase-space integral is factorized into an ordinary
integral over the parallel components and a Lorentz-invariant
three-particle phase-space integral in $2+1$ dimensions
with masses $\mu(\kpa{i})$ and $\mu(\qp)$.

In order to perform the integral over the orthogonal components,
the phase-space integrand $|A|^2$ in \eqref{eq:photon-phase-space-integral}
needs to be rewritten in
a $(2+1)$--dimensional Lorentz-invariant form,
which turns out to be possible.
The explicit expressions are, however, quite complicated. We define the
$(2+1)$-dimensional Lo\-rentz vectors
\begin{subequations}
\begin{align}
\lambda_i^\mu\defeq & \big( \sqroot{\mu_i^2+\koa{i}^2}\,, \;
\vkoa{i} \big) \defeq (\omega_i\,, \vkoa{i}), \\[2mm]
\lambda_q^\mu\defeq &
\big( \sqroot{\mu^2+\qo^2}\,, \; \vqo \big) \defeq  (\omega_q\,, \vqo).
\end{align}
\end{subequations}
All the factors in the
integrand which depend on the
orthogonal momenta can then be written as
products of contractions of  $\lambda_q$ and
$\lambda_i$, where each particle ``momentum''
occurs in second order. Additional factors that only depend on
$\kpa{i}$  or $\omegapa{i}$ are pulled out of the inner integrals.

Next, we perform the Lorentz-invariant phase-space integrals
for the $2+1$ dimensions. These can be solved by mass convolutions
which, unlike for the $(3+1)$-dimensional case,
do not lead to elliptic integrals
but to polynomials in the formal masses $\mu$ and $\mu_i$.

After this integration, we are effectively left with
a double integral over momentum components
parallel to $\widehat\vz$ and an integrand that
is, except for a factor $1/\omegap(\qp)$ and the squared
normalization factors of the polarization vectors, a polynomial in
the momentum components and the parallel energies. Despite this
relatively simple structure, we have not been able to perform the integral
analytically. However, both by numerical methods and by Laurent
expansion (with a low-momentum cutoff as additional approximation),
the behavior for large momentum component $\qpabs$ could be
extracted. The result has been given in Eq.~\eqref{eq:c-numresults}
of the main text.

For a $\plus$ photon at rest ($\vec q = \vec 0$), the
decay parameter of the only open channel $\plus\to\minus\minus\minus$
was expressed in Ref.~\cite{AK:photon-decay} as
\begin{equation}
\gamma_\plus(0) = 2m\;
\frac{K^2\,m^{9}}{3!\; 512\pi^5\; 4\pi^4}\; I\text,\\[3mm]
\end{equation}
with  a preliminary estimate $I\approx 0.2$
for the dimensionless phase-space integral. We have
now obtained an improved numerical value for this integral,
\begin{equation}
I\approx0.02773\text,
\end{equation}
which is lower by a factor $7$ (most likely, the numerical accuracy
of Ref.~\cite{AK:photon-decay} was insufficient to get a reliable
estimate). Writing $\gamma_\plus(0) = c_0\,K^2\,m^{10}$, we then have
$c_0 \approx 1.514\times 10^{-10}$ at $\qp =0$.
For larger momentum components $\qpabs$ and the exclusive decay channel
$\plus\to\minus\minus\minus$, the decay parameter $\gamma_\plus(\qp)$
reaches the power-law behavior \eqref{eq:gamma-photon-splitting} with
constant \eqref{constPtoMMM},   
which holds for $m \ll \qpabs \ll M^2/m$.
\end{appendix}

\end{document}